\documentclass[a4paper,aps,preprintnumbers,twocolumn,amsmath,amssymb,prl,showpacs]{revtex4}
\usepackage{bbm}
\usepackage{mathrsfs}
\usepackage{amsmath}
\usepackage[dvips]{graphicx}
\usepackage{epsfig}
\usepackage[dvips]{graphics,graphicx}

\newcommand{\be}{\begin{equation}}
\newcommand{\ee}{\end{equation}}
\newcommand{\bea}{\begin{eqnarray}}
\newcommand{\eea}{\end{eqnarray}}
\newcommand{\bes}{\begin{split}}
\newcommand{\ees}{\end{split}}

\renewcommand{\vec}[1]{\mathbf{#1}}

\usepackage[dvips]{graphics,graphicx}

\begin{document}
\title{Cooling into the Spin-Nematic State for a Spin-1 Bose gase in an optical lattice}
\author{Ming-Chiang Chung and Sungkit Yip}
\affiliation{Institute of Physics, Academia Sinica, Taipei 11529, Taiwan}

\begin{abstract}
  The possibility of adiabatically cooling  a spin-$1$ polar Bose gas to a spin-nematic phase is theoretically discussed. The relation between the order parameter of  the final spin-nematic phase and the starting temperature of the spinor Bose gas is obtained both using the mean-field approach for the high temperature and spin-wave approach for the low temperature. We find that there exists a good possibility to reach the spin-nematic ordering starting with spinor antiferromagnetic Bose gases.   
\end{abstract}

\pacs{37.10.Jk,03.75.-b,75.25.+z}

\date{\today}
\maketitle

A trapped Bose-Einstein condensate (BEC) loaded in  periodic counter-propagating laser beams to form an optical lattice \cite{Jessen} becomes a very important technique to explore the general properties of condensed-matter physics  because of the flexible tunability of the parameters, for example, changing the atom-atom interaction strength using  Feshbach resonances \cite{Stwalley, Tiesinga} or reducing the hopping constant by increasing  laser intensity. Therefore  the optical lattice can imitate the ion-lattice potential one encounters in solid-state materials without the disadvantage that a real compound has - the impurity. The pureness of an optical lattice enables the exploration of the strong-correlated systems which is the main issue of the condensed-matter physics. 

Recently, a theoretical suggestion about  the possibility to  adiabatically cool a weakly interacting spin-$1/2$ Fermi gas to an antiferromagnetic phase of Hubbard model has been made \cite{Stoof}, however, experimentally it is difficult to be fulfilled due to  the extremely  low temperature one has to reach to cool Fermi gases.  On the other hand, Bose gases with high spins in a lattice offer different rich phases, such as dimerized states \cite{Yip, Fazio} and  nematic \cite{chubukov, troy}  orders.  The increasing interests in spinor optical lattices provide a chance to discuss open problems in the theory of quantum magnetism, which only zero-temperature physics has been so far discussed \cite{Demler}. Especially the maturity of the technique in trapping spinor Bose gases and the large entropy contained in the high spin systems benefit the opportunity to adiabatically cool a spinor Bose gas to an exotic quantum magnetic state.  In this Letter we discuss the finite-temperature properties of spin-nematic states and the possibility of achieving the spin-nematic state  from a spin-$1$ Bose gas. For simplicity, we shall limit ourselves to a sample with no net magnetization.

We start with a spin-$1$ Bose gas with antiferromagnetic interaction, for instance $^{23}Na$ \cite{Ketterle},  which  ground state is a polar state with zero spin expectation value \cite{Ho}.
Now we adiabatically launch  periodic potentials using conter-propagating laser beams to form an isotropic rectangular  optical lattice in 3D. In the case of only one atom per each potential well, Bose-Hubbard model can effectively represent  the mechanism for atoms in an optical lattice \cite{Jaksch}. The hopping constant $t$ decays exponentially with the square root of the laser intensity and the atom-atom interaction strength $U_S$ depends on the total spin $S=0,2$, where $S=1$ is excluded due to the identity of bosons with one atom per each well. In three dimensions, the on-site interaction is effectively a linear function of the s-wave scattering length: $U_S\propto a_S$. With a suitable large atom-atom coupling $ U/t \gg 1$, the Bose-Hubbard model can be replaced by an effective Hamiltonian  
$H =  \sum_{\langle i,j \rangle} H_{ij}$\cite{Yip}, where  
\be \label{EffHam}
  H_{ij} = J({\mathbf S}_i\cdot {\mathbf S}_j) + K ({\mathbf S}_i\cdot {\mathbf S}_j)^2 + J-K , 
\ee
 with $J,K$ defined as $J = -2t^2/U_2$, $K = -2t^2/3U_2-4t^2/3U_0$ and $\langle i,j\rangle$ symbolized as the next-neighbor sites. 
For $^{23}Na$ atoms, for instance,  since $a_2 \approx 52 a_B$ and $a_0 \approx 46 a_B$ with Bohr radius $a_B$, we have $U_2 > U_0 >0$, therefore $K < J < 0$. In this case, $|U_2-U_0|/U_0 \ll 1$, hence $|J| \sim |K| $ and $|J-K| \ll |J|, |K|$.
The parameters can be mapped into a unit circle by defining $\gamma \equiv \tan^{-1} K/J$. 
For  $K<J<0$, $-3\pi/4< \gamma <-\pi/2 $. The two limits depend on the atom-atom coupling: if $|U_2-U_0| \ll U_0 \lesssim U_2$ as for $^{23}Na$, $\gamma \rightarrow -3\pi/4$. On the contrary, if $U_2\gg U_0 >0$, $\gamma \rightarrow -\pi/2$. Since the scattering lengths are tunable using Feshbach resonances, one can explore all parameter regime experimentally.  

A numerical calculation by Troy et. al. \cite{troy}  suggested that the ground state should be a spin-nematic state for an isotropic 3D lattice if $-3\pi/4< \gamma <-\pi/2 $. A nematic state comes from the spontaneously symmetry breaking of $O(3)$ rotation symmetry, though conserving the time-reversal symmetry. Therefore, the expectation value of any spins vanishes: ($\langle S^{\alpha}\rangle = 0, \alpha = x,y,z$) due to the conserved time-reversal symmetry, while the quadrupole orders defined as ${\cal Q}^{\alpha\beta} \equiv 1/2 (\langle S^{\alpha}S^{\beta}+S^{\beta}S^{\alpha}\rangle -4/3\delta^{\alpha\beta})$ have finite expectation values due to the broken  symmetry of the $O(3)$ rotation.  With these two features in mind, one can define a new set of basis: $|x\rangle = \frac{1}{\sqrt{2}}(-|1\rangle + |-1\rangle), |y\rangle = \frac{i}{\sqrt{2}}(|1\rangle + |-1\rangle), |z\rangle = |0\rangle $, where $|m_z=-1,0,1\rangle$ are the eigenstates of  $S^z$. Using the new basis, it can be easily verified that $S^{\alpha} |\beta \rangle = i \epsilon_{\alpha\beta\gamma} |\gamma\rangle$  with $\alpha\beta\gamma = x,y,z$. The spin expectation value of a pure state $|\psi\rangle = \sum d_{\alpha} |\alpha\rangle$ is  given by $\langle \psi | S^{\alpha} |\psi \rangle = i \epsilon_{\alpha\beta\gamma} d_{\beta} d_{\gamma}^{\star}$, hence, if $d_{\alpha}$ are all real, the spin expectation values vanish. Therefore,
for nematic states, there exists a vector ${\mathbf d} = (d_x,d_y,d_z)$ that breaks $O(3)$ symmetry. 
Without lost of generality, one can rotate the system to make the vector ${\mathbf d}$ as $z$ direction. In this way, the five possible components of quadrupole moments with the normality constraint ($d_x^2+d_y^2+d_z^2=1$) for a spin-1 system can be reduced to a single component: $d_z^2-1/3$. 

The same argument  described above for pure states can be applied to mix states.       
A density matrix of a single site is defined as $\hat{\rho} = \sum_{\alpha,\beta=x,y,z} \rho^{\alpha\beta} |\alpha\rangle \langle \beta |$. The spin expectation values are thus given by 
\be
  \langle S^{\alpha} \rangle = i \epsilon_{\alpha\beta\gamma} \rho_{\beta\gamma} = \mp 2 {\mbox{Im}} \rho_{\beta\gamma}.
\ee
The spin ordering is therefore associated with the imaginary part of $\hat{\rho}$. Since $\hat{\rho}$ is hermitian, the imaginary part of a density matrix  denotes the antisymmetry part of it. If the density matrix is real and hence symmetric, the spin expectation values disappear, and the system conserves the time-reversal symmetry.  A real symmetric matrix can always be diagonalized, thus with a suitable choice of axes: 
\be \label{rho}
 \hat{\rho} = \rho^{xx} |x\rangle\langle x|  + \rho^{yy} |y\rangle\langle y|  +  \rho^{zz} |z\rangle\langle z|. \ee 
 If, as in the $T=0$ state,  the broken $O(3)$ symmetry is in the $z$ direction, one has $\rho^{xx}=\rho^{yy}
\neq \rho^{zz}$.  The expectation values of the spin square thus have the form: $\langle (S^{x})^2 \rangle = \langle (S^{y})^2 \rangle \neq \langle (S^{z})^2 \rangle$. Using this broken symmetry, the only nonzero element of the five quadrupole order parameters: $S^{\alpha}S^{\beta}+ S^{\alpha}S^{\beta} (\alpha \neq \beta)$, $(S^{x})^2-(S^{y})^2$ and $(S^{z})^2-2/3$ is the expectation value in $z$ direction: 
\be \label{ExpVofS}
(S^{z})^2-\frac{2}{3} = \frac{1}{3} (\rho^{xx}+\rho^{yy}) - \frac{2}{3} \rho^{zz} = -(\rho^{zz}-\frac{1}{3}).
\ee
Here we have used the normality condition: ${\mbox Tr} \hat{\rho} =1$. With the definition: $q \equiv \rho^{zz}-\frac{1}{3}$, the nematic system only depends on the parameter $q$, the normal vectors ${\mathbf d}$ and $-{\mathbf d}$ defined as  unit vectors in $\pm z$ direction. 

The two-body Hamiltonian (\ref{EffHam}) can be rewriten as $H_{ij} = e_0 P_{ij}^{(0)} + e_2 P_{ij}^{(2)}$, where $e_0 = -4t^2/U_0$, $e_2 = -4 t^2/U_2$ and $P_{ij}^{(S)}$  is the projection operator which projects the pair $i$ and $j$ into a total hyperfine spin $S$  state. Since $P_{ij}^{(0)}$ projects the states to $|S=0\rangle$, the operator should be rotational invariant. From this symmetry observation, the only possible rotational invariant state is $\frac{1}{\sqrt{3}}  \sum_{\alpha=x,y,z} |\alpha\rangle_{i} |\alpha\rangle_{j} $ which has to be proportional to $|S=0\rangle$. Therefore 
\be \label{zeroProj}
P_{ij}^{(0)}= \frac{1}{3}\sum_{\alpha.\beta=x,,z}  |\alpha\rangle_{i} |\alpha\rangle_{j} {_i}\langle\beta|{_j}\langle\beta|. \ee
 As for $P_{ij}^{(2)}$, one needs five orthonormal states which are also orthogonal to $|S=0\rangle$. Since the states corresponding to $S=2$ are symmetric under the coordinate exchange $i \leftrightarrow j$, the five orthonormal states can be written down by constructing the second order polynomials corresponding to a symmetric rank-2 tensor. One of the possible construction is $|I_{\alpha\beta}\rangle = \frac{1}{\sqrt{2}}  (|\alpha\rangle_{i} |\beta\rangle_{j} +  |\beta\rangle_{i} |\alpha\rangle_{j} $) for $\alpha \neq \beta$, $|I_{0}\rangle = \sqrt{\frac{2}{3}} (|z\rangle_{i} |z\rangle_{j} -\frac{1}{2} |x\rangle_{i} |x\rangle_{j}  -\frac{1}{2} |y\rangle_{i} |y\rangle_{j}) $ and $ |I_{1}\rangle = \frac{1}{\sqrt{2}}  (|x\rangle_{i} |x\rangle_{j} - |y\rangle_{i} |y\rangle_{j})$. Therefore 
\be \label{twoProj}
P_{ij}^{(2)} = \sum_{\{I\}} |I\rangle \langle I|.
\ee

\begin{figure}
\center
\includegraphics[width=7.0cm]{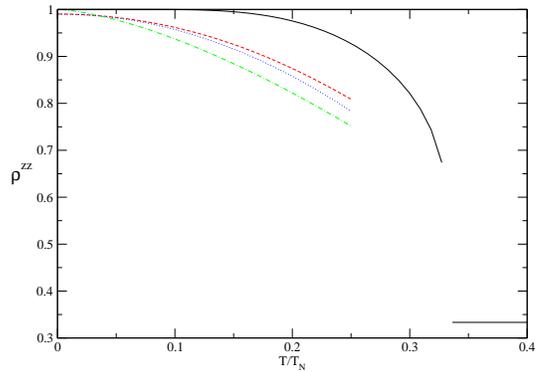}
\caption{(Color online) Order parameter of the nematic state vs. temperature in unit of $k_B T_N=6|K|$.  Solid line is the mean-field result, while the red dashed line represents the spin-wave approach. The two approximations using linear and quadratic dispersion are shown in  blue dotted line and  green dotted-dashed line, respectively.} \label{fig1}
\end{figure}

   The principle of the mean-field approximation is that one reduces a many-body problem to a one-body problem by singling out  one  site and averaging the other sites served as a mean field. A mean-field treatment for a spin-$1$ Bose system has also been done by Chen and Levy\cite{Chen}.  The effective Hamiltonian is thus a single site operator given  by 
\be
    H_{eff} = z {\mbox{Tr}}_j [ \hat{\rho}_j H_{ij} ]  = z {\mbox{Tr}}_j \left[ \hat{\rho}_j (e_0 P_{ij}^{(0)} + e_2 P_{ij}^{(2)})\right], 
\ee   
with the coordinate number $z$. For a rectangular three-dimensional lattice, $z=6$. Using Eqs.(\ref{rho}), (\ref{zeroProj}) and (\ref{twoProj}) one can obtain the effective Hamiltonian as 
\be
  H_{eff} = z \sum_{\alpha = x,y,z} (K \rho^{\alpha\alpha} + \frac{e_2}{2}) |\alpha\rangle \langle \alpha|.  
\ee  
The self-consistent equation for the only nonzero quadrupole order $q \equiv \rho^{zz} - \frac{1}{3}$ which comes from the relation $\rho^{zz} = e^{- H_{eff}/k_BT}/{\mbox{Tr}} e^{- H_{eff}/k_BT}$ reads
\be \label{MeanField}
   \frac {3q}{2}  = \frac{\exp{(\frac{z|K|}{k_B T}\frac{3q}{2})}-1}{\exp{(\frac{z|K|}{k_B T}\frac{3q}{2})}+2}. 
\ee 
$q=0$ is always a solution, actually it represents the disordered phase $\rho^{xx}=\rho^{yy} = \rho^{zz} = 1/3$, correspondingly, the number of particle per site $n_{m_z}$ in the state $|m_z=-1,0,1\rangle$ obeys $n_0=n_{\pm 1}$. This state conserves all symmetries and gives the equal spin expectation values: $\langle (S^{x})^2 \rangle = \langle (S^{y})^2 \rangle = \langle (S^{z})^2 \rangle = \langle S^2 \rangle/3 = 2/3 $ as the result indicated by Eq. (\ref{ExpVofS}).

The solid line in Fig.~\ref{fig1} shows the mean-field result for the order parameter $\rho^{zz}$ as a function of  $T/T_N$ where $T_N=6|K|$. In this figure we found a first order phase transition with the transition temperature $T_c/T_N \approx 0.36$. Below $T_c$, the system has the broken $O(3)$ symmetry in the $z$ direction, therefore $\rho^{zz} > 1/3$. Above $T_c$, $\rho^{zz} = 1/3$ for all temperaure. The order parameter jumps discontinuously from the ordered state to the disordered state at the transition temperature $T_c$. We have to mention that the transition temperature predicted by the mean-field theory is not correct as the case for the Ising model \cite{Huang}, and the order parameter in the low temperature decreases exponentially, as shown in Fig~\ref{fig1}, which is an artifact of the mean-field approach. However, the mean-field solution gives us a qualitative description about the ordered-disordered phase transition. 

 The entropy per unit site  given by the definition 
$s = -k_B \mbox{Tr} \hat{\rho} \ln \hat{\rho} $ reads:
 \be \label{Entropy} 
s= -k_B \left[\rho^{zz} \ln\rho^{zz} + (1-\rho^{zz})\ln\left(\frac{1-\rho^{zz}}{2}\right) \right].\ee 
The critical order parameter $\rho_c^{zz} = 0.67$ at $T_c$, which gives the critical entropy: $s_c = 0.863$ according to the Eq.(\ref{Entropy}). The value of the critical entropy is quite large due to the large entropy capacity for spin-$1$ systems. Above $T_c$, the entropy jumps to a constant: $s_d=k_B \ln{3}$. The number $3$  indicates the mean-field solution takes equal probability for each conserved $O(3)$ symmetry for all temperatures $T>T_c$.       

\begin{figure}
\center
\includegraphics[width=7.0cm]{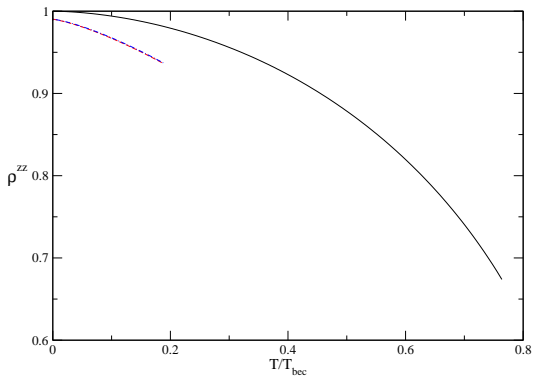}
\caption{(Color online) Order parameter of the nematic state vs. starting temperature of spinor Bose gases in unit of $T_{\mbox{bec}}$ using the mean-field approach (solid line) and the spin-wave theory (red dashed line). For the red dashed line, the interacting spectrum for the initial Bose gas is used, while the blue dashed-dot line shows the result with an ideal Bose gas as an initial state. } \label{fig2}
\end{figure} 

At low temperature, the mean-field theory fails because it ignores long-wavelength fluctuations. 
This problem has been solved using spin-wave theory by Chubukov \cite{chubukov}
by approximating the spin-up state $|1\rangle$ and the spin-down state
$|-1\rangle$ as two bosonic states and $|0\rangle$ as a vacuum state. This approximation works very well if the excitation number per each site is less than one.  The resulting energy spectrum consists two degenerate branches 
\be \label{EnergyNem}
   \varepsilon_k = (- z J ) (1+\delta)^{1/2}\left\{ (1-\nu_k)[1-\nu_k + \delta(1+\nu_k)]\right\}^{1/2}, 
\ee  
where $\delta = \tan{\gamma}-1$ and the tight binding function $\nu_k \equiv 1/z \sum_{\vec{r_{\eta}}} e^{i \vec{k}\cdot\vec{r_{\eta}}}$ with the all next-neighbor vectors $\vec{r_{\eta}}$. In the long-wavelength limit and for $\delta \ll 1$, the energy spectrum is linear: $\varepsilon_k = ck$  where $c = -zJ\sqrt{2\delta/z} a$ with the lattice spacing $a$. Therefore these two energy branches correspond to two Goldstone modes due to the two remaining symmetries for a broken $O(3)$ symmetry.

The order parameter $\rho^{zz}$ corresponds to the density still remaining in the vacuum state (remember $ |z\rangle = |0\rangle$). In the momentum space, $\rho^{zz}$ can be separated into two terms: zero temperature  depletion $\eta(\delta)$ and thermal depletion $\eta_T(\delta, T)$: $\rho^{zz} = 1 - \eta_0(\delta) - \eta_T(\delta, T) $, where 
\bea
    \eta_0(\delta) & = & a^3 \int \frac{\mbox{d}^3 k}{(2\pi)^3} \left\{\frac{\delta+(1-\nu_k)}{\sqrt{(1-\nu_k)[2\delta+(1-\nu_k)]}}-1\right\} \nonumber \\
     \eta_T(\delta, T) & = &   2 a^3 \int \frac{\mbox{d}^3 k}{(2\pi)^3} \frac{\delta+(1-\nu_k)}{\sqrt{(1-\nu_k)[2\delta+(1-\nu_k)]}}  \nonumber \\
      & & \times \frac{1}{(e^{\frac{\varepsilon_k}{k_B T}}-1)}. 
\eea         
The red dashed line in Fig~{\ref{fig1}} shows  $\rho^{zz}$ obtained by using the spin-wave approach with the parameters $K/J = 1.1  (\delta=0.1)$, which are very close to the parameters $K/J =1.089$ for $^{23}Na$ . Contrary to the result obtained from the mean field, for which $\rho^{zz} = 1 $ at zero temperature representing the classical nematic ground state, the quantum fluctuation of the spin wave gives $\eta_0 = 0.009$.   In Fig.\ref{fig1} two approximated curves are also shown.  In the  regime $T/T_N \ll \delta$, where the energy dispersion is linear, $\eta_T = \sqrt{3}(1+\delta)^2/(2\sqrt{\delta}) (T/T_N)^2$, as shown with the blue dotted line. 
    Beyond the linear regime, for $ \delta \ll T/T_N \ll 1$, $\varepsilon_k \approx |J| k^2 a^2$. In this case the zero temperature fluctuation is zero and $\eta_T = 2[3(1+\delta)/2\pi]^{3/2} \zeta(3/2) (T/T_N)^{3/2}$ as shown with green dotted-dashed line. We can see that the order parameter of spin wave theory coincides with the  result using linear dispersion at the temperature $T/T_N < 0.1$ and then goes parallel to the approximation with a quadratic dispersion. At higher temperature the spin-wave theory overestimates the depletion to the excited states and hence the order parameter decreases faster than the mean field approach. This is because at higher temperature, more than one boson can occupy one site. This feature can not explain the spin statistics more, therefore the spin-wave theory fails for the high temperature regime.     


 The initial spinor Bose gas has three modes of excitations\cite{Ho}: one corresponds to density excitations with the energy dispersion $\epsilon_0 = [(k^2/2m)^2+ nc_0 k^2]^{1/2}$, and the other two correspond to two spin modes: $\epsilon_2= [(k^2/2m)^2+ nc_2 k^2]^{1/2}$ with the mass of atoms $m$ and the density of condensate $n$. The interaction strengths $c_0$ and $c_2$ is defined as follows: $c_0 = 4 \pi (a_0+2a_2)/3m$ and $c_2 = 4\pi (a_2-a_0)/3m$. If one  starts with a Bose gas at the temperature $ k_BT \gg n c_0 $ for which the interaction is not important, the system is approximated as an  ideal Bose gas. The adiabatic cooling ensures that the beginning state (Bose gas) and the final state (nematics) should have the same entropy
\be 
    s = k_B \frac{5 \zeta(5/2) }{2 \zeta(3/2)} \left(\frac{T}{T_{\mbox{bec}}} \right)^{3/2},  
\ee
where $T_{\mbox{bec}} = 2\pi n^{2/3}/[3\zeta(3/2)]^{2/3}m $, $\zeta(x) \equiv \sum_{n=1}^{\infty} 1/n^x$ is the Riemann zeta function and $s$ is given by Eq.(\ref{Entropy}). The solid line in Fig.~\ref{fig2} shows the mean-field order parameter of the cooled nematic state versus the starting temperature $T/T_{\mbox{bec}}$ of the Bose gas using the mean-field approach. There exits a maximum temperature for the starting state such that $T/T_{\mbox{bec}} < 0.769$. On the other hand, for $s_c< s < s_d$, $0.679 < T/T_{\mbox{bec}} < 0.9$. In this initial temperature regime, the system is cooled to a coexistence of the  ordered  and disordered phase with phase separation due to the first order phase transition. For $T/T_{\mbox{bec}} > 0.9$, the nematic phase can therefore not be reached.

For low temperature, the entropy of bosonic systems has the form   
\be S= k_B \sum_{k,\sigma} \left[\frac{E_{k}^{\sigma} e^{\frac{E_{k}^{\sigma}}{k_B T}}}{k_B T(e^{\frac{E_{k}^{\sigma}}{k_B T}}-1)} - \ln(e^{\frac{E_{k}^{\sigma}}{k_B T}}-1)\right],\ee
where $\sigma$ represents different Goldstone modes and 
$E_k^{\sigma}$ is the corresponding excitation energy with momentum $k$. For the nematic state, $E_k^{\sigma} = \varepsilon_k$ for two degenerate  modes, while for the initial spinor Bose gases  $E_k^{\sigma} = \epsilon_0$ for one density mode and  $E_k^{\sigma} = \epsilon_2$ for two spin modes. By mapping the initial BEC temperature to the cooled nematic temperature with equal entropy, we found the order parameter as a function of the initial BEC temperature as the red dashed line shown in Fig~\ref{fig2}.  At low temperature, the order parameter decays much sooner than that for the mean field, and the zero temperature fluctuation exists. For comparison we also show the order parameter initially with an ideal Bose gas as the blue dotted-dashed line in Fig.~\ref{fig2}.   
 There almost exists no difference for $\rho^{zz}$ between the interacting spinor Bose gas and the ideal Bose gas  as an initial state because the chemical potential $n c_0 $ for the interacting gas is small and hence is easily smeared  out by the temperature.                        

 Moreover, we consider a BEC trapped in a harmonic trap with a trap frequency $\omega$ due to the fact that in reality a BEC of ultracold atoms only exists in a magnetic or optical trap. The nematic exists with uniform $K$ and $J$ in the case that the chemical potential at the center of the trap is less than the atom-atom interaction strength $U_S$ and the entropy at the edges is not dominant. The second condition can be possibly fulfilled because the nematic state carries a lot of entropy. The starting entropy for the trapped gas has the form:
\be
   s = \frac{4\zeta(4)}{\zeta(3)}\left(\frac{T}{T_{\mbox{bec}}} \right)^3,
\ee
where $T_{\mbox{bec}} = n^{1/3}\omega /[3\zeta(3)]^{1/3}$. The starting temperature regime for cooling into a nematic state is $T/T_{\mbox{bec}} < 0.62$ if $s \leq s_c$ is used. Therefore the final result  is similar to Fig.~\ref{fig2}  with a squeezed temperature regime. 

$\rho^{\alpha\alpha}$ can be directly measured since it is the number of particles per site in the $\hat{\alpha}\cdot \vec{S} = 0 $ state where the quantization axis is chosen along the $\alpha$-direction. The ordered state is characterized by $ \rho^{zz} > 1/3 > \rho^{xx(yy)}$. For quantization along $z$-direction, the ordered state can also be identified by $n_0 >1/3 > n_{\pm 1}$. On the contrary, the disordered state has isotropic densities.      However, in reality, there also always exists a small magnetic field which breaks the rotation symmetry due to the quadratic Zeeman effect \cite{Ketterle}. Therefore for $T>T_c$,  $\rho^{zz}$ is slightly larger than $1/3$, so it is not entirely disordered. However, for a small enough quadratic Zeeman field the first order phase transition still exists.     

In summary, we have shown, by means of both mean-field  and  spin-wave approaches,  thermodynamically the order parameter of the nematic state for a spin-1 Bose gas in an optical lattice. With the mean-field approach, we found the first-order transition temperature.   The entropy containing in a nematic state is very large as a result of higher spin systems. Using the spin-wave approach, we found the zero-temperature fluctuation and power-law decay of the order parameter. We furthermore discussed the possibility of cooling a spinor polar Bose gas to a spin-nematic magnetic state. We found that there always exists a large  temperature range under which the spin-nematic state can be reached. 

This research was supported by the National Science Council of Taiwan under Grant No. NSC91-2112-M-001-063.

\bigskip


\end{document}